\documentclass[namedreferences]{SolarPhysics}
\usepackage[optionalrh]{spr-sola-addons} 
\usepackage{graphicx}                    
\usepackage{color}                       
\usepackage{url}                         

\begin{document}

%
%
\newcommand{\br}{{\mathbf{r}}}
\newcommand{\taup}{{\tau_{\rm p}}}
\newcommand{\taug}{{\tau_{\rm g}}}
\newcommand{\dx}{dx}
\newcommand{\dt}{dt}
\newcommand{\id}{\mbox{$\rm d$}} 
\newcommand{\vel}{\mbox{{\boldmath$v$}}} 

\begin{article}

\begin{opening}

\title{Helioseismic Travel-Time Definitions and Sensitivity to Horizontal Flows Obtained From Simulations of Solar Convection}

\author{S\'ebastien~\surname{Couvidat}$^{1}$\sep
        Aaron C.~\surname{Birch}$^{2}$}

\institute{$^{1}$ W.W. Hansen Experimental Physics Laboratory\\
           491 S Service Road, Stanford University\\
           Stanford, CA 94305-4085, USA\\
           email: \url{couvidat@stanford.edu}\\
           $^{2}$ NorthWest Research Associates, CoRA Division\\
	   3380 Mitchell Lane, Boulder, CO 80301, USA\\
           email: \url{aaronb@cora.nwra.com}}
\date{Received: 4 March 2009/ Accepted:}
\runningauthor{S. Couvidat and A.C. Birch}
\runningtitle{Helioseismic Travel-Time Definitions and Sensitivity to Horizontal Flows}

\begin{abstract}

We study the sensitivity of wave travel times to steady and spatially homogeneous horizontal flows added to a realistic simulation of the solar convection performed by Robert F. Stein, Ake Nordlund, Dali Georgobiani, and David Benson.
Three commonly used definitions of travel times are compared. We show that the relationship between travel-time difference and flow amplitude exhibits a non-linearity depending on the travel distance, the travel-time definition considered, and the details of the time-distance analysis (in particular, the impact of the phase-speed filter width). For times measured using a Gabor wavelet fit, the travel-time differences become nonlinear in the flow strength for flows of about $300$ m s$^{-1}$, and this non-linearity reaches almost $60\%$ at $1200$ m s$^{-1}$ (relative difference between actual travel time and expected time for a linear behaviour). We show that for travel distances greater than about 17~Mm, the ray approximation predicts the sensitivity of travel-time shifts to uniform flows.  For smaller distances, the ray approximation can be inaccurate by more than a factor of three.
\end{abstract}


\end{opening}

\section{Introduction}

Since the introduction of time-distance helioseismology (Duvall {\it et al.}, 1993) this method has been applied to investigate various subsurface physical properties of the quiet Sun ({\it e.g.} Kosovichev and Duvall, 1997), especially supergranulation ({\it e.g.} Duvall and Gizon, 2000; Zhao and Kosovichev, 2003; Hirzberger {\it et al.}, 2007), and sunspots ({\it e.g.} Kosovichev, Duvall, and Scherrer, 2000; Zhao, Kosovichev, and Duvall, 2001; Couvidat, Birch, and Kosovichev, 2006). In these studies, the main emphasis has been on mapping both the subsurface material flows and deviations of the sound speed from a reference solar model.
Travel times of wavepackets (whether from acoustic or surface gravity waves) are the basic measured quantities that are then inverted to produce estimates of the flow velocity and sound-speed perturbations.
Travel times are measured from temporal cross-covariances. Over the years different definitions of these travel times have been employed ({\it e.g.}, Kosovichev and Duvall, 1997; Gizon and Birch, 2002; Gizon and Birch, 2004). For a given cross-covariance, these definitions, in general, yield different travel times. It is therefore important to understand how these definitions perform in simple situations where the subsurface conditions are known.
In this paper we study the sensitivity of these definitions to horizontal, steady, and uniform flows. We use vertical velocity data from a realistic numerical simulation of solar convection (Benson, Stein, and Nordlund, 2006). We introduce steady and spatially uniform horizontal flows to this vertical velocity data cube and measure the perturbations in the travel times induced by these flows. We also compare the observed relationship between travel-time differences and flow velocities with the relationship predicted by ray approximation kernels (Kosovichev and Duvall, 1997) commonly used for the inversions.
We limit ourselves to the study of acoustic modes ({\it p} modes), neglecting the surface-gravity modes ({\it f} modes).
The {\it f}-mode case was studied in detail by Jackiewicz {\it et al.} (2007).
In Section~2 we briefly present the numerical simulation that we used for this work. In Section~3 we remind the reader of the fundamentals of the time-distance formalism. In Section~4 we present our results, and we conclude in Section~5.

\section{Numerical Simulation}

We work with simulated line-of-sight velocity data [$\phi(\mathbf{r},t)$ where $\mathbf{r}$ is the horizontal position vector and $t$ is time] obtained from a numerical three-dimensional simulation of convection in the upper solar convection zone. The numerical slab is 96 Mm$\times$96 Mm wide and 20 Mm deep. The spatial resolution is 96 km horizontally and varies from 12 to 75 km vertically. The temporal resolution is 10 seconds and we use 8.5~hours of simulated data. 
This simulation was performed by Robert Stein, Ake Nordlund, Dali Georgobiani and David Benson, and is publicly available on the website \url{sha.stanford.edu}.
A very similar simulation (48 Mm$\times$48 Mm horizontally) has already been extensively analyzed by Georgobiani {\it et al.} (2007), Braun {\it et al.} (2007), and Zhao {\it et al.} (2007). The latter applied the time-distance formalism to these data to study, among other things, the validity of the inversion procedures. A detailed presentation of the code that produced these data can be found in Benson, Stein, and Nordlund (2006) and Stein and Nordlund (2000).
Here we only work with the vertical velocity at the surface of the numerical simulation. The data cube we use is rebinned four times in the horizontal direction (yielding a final spatial resolution $\dx$=384 km) and six times in time (final temporal resolution $\dt$=60 seconds).
One advantage of using this numerical simulation in place of solar data is that horizontally the boundary conditions are periodic. Therefore it is possible to simulate the impact of a steady horizontally uniform flow by simply shifting the vertical velocity maps in the horizontal direction at each time-step, using the shift theorem in the Fourier domain.
Simulating an advection flow on actual solar data, {\it e.g.} on data from the SOI/MDI instrument (Scherrer {\it et al.}, 1995), is possible (Jackiewicz {\it et al.}, 2007) but more complicated and under normal conditions the simulated flow is non-uniform in latitude.
The power spectrum of the vertical velocity $\phi(\mathbf{r},t)$ from the numerical simulation is very similar to the actual solar oscillations power spectrum as determined by SOI/MDI (Georgiobani {\it et al.}, 2007). Therefore we expect that the results we obtain here should be fairly representative of the actual solar case.

\section{Time-Distance Formalism: A Brief Overview}

\subsection{Cross-Covariances}

We use the time-distance measurement procedure described in Couvidat, Birch, and Kosovichev (2006), except that we work with a center-quadrant geometry instead of a center-annulus one. We only deal with acoustic waves ({\it p} modes) and filter out the surface gravity waves ({\it f} modes) (see Jackiewicz {\it et al.}, 2007, for a recent study of the travel times of {\it f} modes). 
The measurement of the travel times begins with the computation of cross-covariances between two observation points $\br_1$ and $\br_2$. To reduce the noise level of the cross-covariances,
the data cubes are phase-speed filtered (Duvall  {\it et \mbox{al.}}, 1997). The data cube $\phi(\mathbf{r},t)$ is filtered in  the  Fourier  domain using  a Gaussian filter $F(k,\omega;\Delta)$, where $k$ is the horizontal wavenumber and $\omega$ is the temporal angular frequency,  
for each travel distance $\Delta=\| \br_2-\br_1 \|$. An {\it f}-mode filter is also applied to remove the {\it f}-mode ridge.
The point-to-point cross-covariances are averaged 
over quadrants (arc of 90$^\circ$) of radius 
$\Delta$ centered on the point $\br_1$. The quadrant geometry (two separate pairs of quadrants: East\,--\,West and North\,--\,South) is used to retrieve the subsurface horizontal flow velocity. Such point-to-quadrant
cross-covariances are  computed for four central distances ($\Delta$) ranging from $6.2$ Mm to $16.95$ Mm. We do not select larger $\Delta$ because there is a small amount of wave reflection at the bottom boundary of the numerical simulation (Zhao {\it et al.}, 2007) that seems to alter the shape of the cross-covariances. The larger the $\Delta$ the deeper the wavepackets reach, and the more they are affected by these reflected waves. Ray theory shows that for $\Delta=16.95$ Mm, the lower turning point of the wavepackets is located at a depth of about $4.5$ Mm, far from the bottom boundary.
To further increase the signal-to-noise ratio of the cross-covariances, we average them in groups of five distances ($\Delta$) for each central distance.
For instance, the final cross-covariances at the central distance $\Delta=6.2$ Mm are averages of five cross-covariances computed at distances ranging from $\Delta=3.7$ to $8.7$ Mm (see details in Couvidat, Birch, and Kosovichev, 2006). 
The values of $\Delta$ and properties of the phase-speed filters used in this paper are summarized in Table \ref{table.1}. 

\begin{table}[t]
\caption{Parameters for the time-distance analysis: distances $\Delta=\| \br_2 -\br_1\|$, range of $\Delta$, properties of the Gaussian phase-speed filters applied to the Fourier transform of the vertical-velocity data cube ($v_0$ is the central phase speed and FWHM is the full width at half maximum), centers of the time window used to measure the travel times, and selected phase travel times ($\taup$) of the reference cross-covariances $C^{\mathrm{ref}}(\Delta,t)$.}
\label{table.1}
\begin{tabular}{cccccc}
\hline
$\Delta$ (Mm) & range (Mm) & $v_0$ (km s$^{-1}$) & FWHM (km s$^{-1}$) & $t_0$ (min) & $\taup$ (min) \\[3pt]
\hline
6.2  & 3.7-8.7  &12.77 & 6.18 & 19.0 & 10.48\\
8.7  & 6.2-11.2 &14.87 & 6.18 & 21.4 & 12.85\\
11.6 & 8.7-14.5 &17.49 & 6.18 & 24.4 & 19.65\\     
16.95& 14.5-19.4&24.82 & 9.09 & 28.7 & 29.05\\
\hline
\end{tabular}
\end{table}

\subsection{Travel-Time Definitions}

The point-to-quadrant cross-covariances  are  used to determine the travel times of the acoustic wavepackets.
Traditionally, cross-covariances have been fit with a Gabor wavelet defined as (Kosovichev and Duvall, 1997):
\begin{equation}
G(A,\omega_0,\delta\omega,\taup,\taug;t) = A \,\, \cos\left\{\omega_0(t-\taup)\right\} \,\, \exp\left(-\frac{\delta\omega^2}{4}(t-\taug)^2 \right) \label{eq.1}
\end{equation}
where $A$ is the amplitude of the wavelet, $\omega_0$ is the central temporal frequency of the wavepacket, $\delta\omega$ is a measure of its frequency width, $\taup$ is the phase travel time, and $\taug$ is the group travel time.
It is customary to use only phase travel times with the time-distance formalism, because group travel times are more difficult to retrieve accurately (the center of the Gaussian envelope is usually poorly estimated) and therefore group travel-time maps are significantly noisier than phase travel-time maps.

Recently, two more definitions of the wavepacket travel times based on studies in geophysics were added. Gizon and Birch (2002) [hereafter GB02] define the travel times $\tau_\pm(\br,\Delta)$ as:
\begin{equation}
\tau_\pm(\br,\Delta)=\mathrm{arg \, min}_t{X_\pm(\br,\Delta,t)}
\end{equation}
where $X_\pm(\br,\Delta,t)$ are the following functions:
\begin{equation}
X_\pm(\br,\Delta,t)=\dt \sum_{t'} f(\pm t') [C(\mathbf{r},\Delta,t')-C^{\mathrm{ref}}(\Delta,t'\mp t)]^2
\end{equation}
where $f(t)$ is a one-sided window function that selects the first-bounce skip of the cross-covariance (the skip corresponding to waves travelling directly from the center to the quadrant) on the positive-time or negative-time branches of the cross-covariance. The same window function is also used when fitting the cross-covariance by a Gabor wavelet even though it does not appear explicitly in Equation (\ref{eq.1}). The function $C^{\mathrm{ref}}(\Delta,t)$ is the reference cross-covariance.
Usually $C^{\mathrm{ref}}(\Delta,t)$ is either obtained from a model of the solar oscillation power spectrum, or is a spatial average of the measured cross-covariances over a region of the quiet Sun.

Lastly, Gizon and Birch (2004) [hereafter GB04] proposed another definition of the travel time that is linear with respect to the cross-covariance:
\begin{equation}
\tau_\pm(\br,\Delta)=\dt \, \sum_t W_\pm(\Delta,t) \, [C(\br,\Delta,t)-C^{\mathrm{ref}}(\br,\Delta,t)]
\end{equation}
where the $W_\pm$ are weight functions:
\begin{equation}
W_\pm(\Delta,t)=\frac{\mp f(\pm t) \, \dot C^{\mathrm{ref}}(\Delta,t)}{\dt \, \sum_{t'} f(\pm t') \, [\dot C^{\mathrm{ref}}(\Delta,t')]^2}
\end{equation}
and $\dot C^{\mathrm{ref}}(\Delta,t)$ denotes the time derivative of the reference cross-covariance.

As emphasized by the use of the $\pm$ symbol, for each of the three definitions of travel times we can measure $\tau$ on the positive-time branch (for wavepackets travelling in a specific direction, {\it e.g.} from the South to the North) and on the negative-time branch (for wavepackets travelling in the opposite direction).
Therefore we  measure four different kinds of travel time:
$\tau_{\mathrm{W}}(\br,\Delta)$ is the travel time of wavepackets traveling westward over a horizontal distance $\Delta$ centered on $\br$; $\tau_{\mathrm{E}}(\br,\Delta)$ is for eastward traveling waves; $\tau_{\mathrm{S}}(\br,\Delta)$ is for southward traveling waves; and finally $\tau_{\mathrm{N}}(\br,\Delta)$ is for northward traveling waves. The corresponding travel-time perturbations are produced by subtracting the travel time obtained from the reference cross-covariance. For instance,  $\delta\tau_{\mathrm{N}}(\br,\Delta)=\tau_{\mathrm{N}}(\br,\Delta)-\tau_{\mathrm{N;ref}}(\Delta)$ where $\tau_{\mathrm{N;ref}}(\Delta)$ is the travel time of the reference cross-covariance. Here, $C^{\mathrm{ref}}(\Delta,t)$ is obtained by averaging over $\br$ the center-to-annulus cross-covariance.
In this paper we focus on $\delta\tau_{\mathrm{NS}}(\br,\Delta)=\delta \tau_{\mathrm{N}}(\br,\Delta)-\delta \tau_{\mathrm{S}}(\br,\Delta)$, the North\,--\,South difference travel-time perturbation, that is, in a first approximation, only sensitive to horizontal flows in the North\,--\,South direction. With this convention a northward flow will result in a negative $\delta\tau_{\mathrm{NS}}$.

\section{Results}

We added southward flows to the velocity data cube from the numerical simulation by applying the shift theorem in the Fourier domain. At each time step ($t$) we shift the velocity maps $\phi(\mathbf{r},t)$ in the $y$ direction (from North to South) by the amount $\delta y= u \, t$ where $u$ is the amplitude of the imposed flow.
Again, the advantage of using the simulation for this purpose is that its boundary conditions are periodic.
We tested 12 flow amplitudes: $u=0$, $100$, $200$, $300$, $400$, $500$, $600$, $700$, $800$, $900$, $1000$, and $1200$ m s$^{-1}$.
Even in the absence of any additional flow (case $u=0$ m s$^{-1}$) the simulation already includes a weak northward flow. This flow is horizontally non-uniform and its velocity varies slightly with depth, with a maximum amplitude of the order of $35$ m s$^{-1}$. The simulation also includes a horizontally non-uniform eastward flow that is somewhat stronger (with an average velocity of about 150 m s$^{-1}$ in the first few Mm underneath the surface) and is strongly depth dependent. The presence of this stronger eastward flow explains why we favor the North\,--\,South direction in the results presented here. However we also tested the addition of eastward flows and their impact on the East\,--\,West travel-time difference, and obtained similar results.
In the rest of this paper we use two kinds of Gaussian phase-speed filters when computing the cross-covariances: those whose parameters are listed in Table~\ref{table.1} (hereafter standard phase-speed filters), and those whose FWHM was multiplied by four (hereafter broad phase-speed filters).

\subsection{Impact of the Horizontal Flows on the Cross-Covariances}

\begin{figure}[ht]
\centering
\includegraphics[height=160mm]{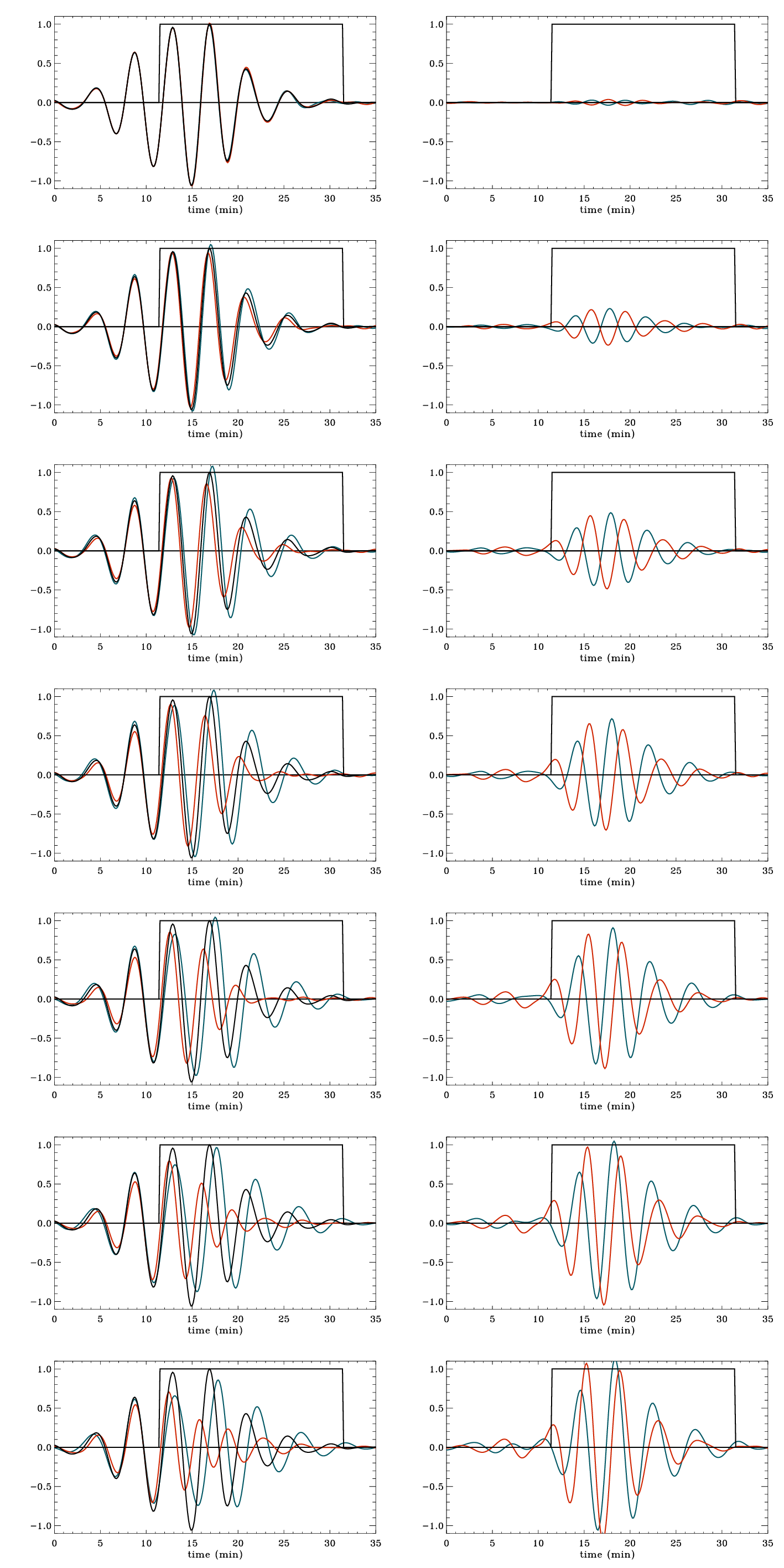}
\caption{Left column: spatially-averaged northward (blue) and southward (red) cross-covariances for different southward flow velocities, at distance $\Delta=8.7$~Mm, and for standard phase-speed filters. The flow amplitudes are, from upper to lower panels: 0, 200, 400, 600, 800, 1000, and 1200 m s$^{-1}$. In black is the reference cross-covariance $C^{\mathrm{ref}}(\Delta,t)$. The temporal window applied when measuring the wavepacket travel times is also shown. Right column: the difference between the northward (blue) or southward (red) cross-covariance and $C^{\mathrm{ref}}(\Delta,t)$.}
\label{fig.0} 
\end{figure}

The impact of the additional uniform horizontal flows on the spatially averaged cross-covariances is shown in Figure~\ref{fig.0} (at $\Delta=8.7$~Mm).
Because the flows are southward, the spatially averaged cross-covariance for the northward-going wavepackets (hereafter northward cross-covariance) is shifted toward larger times compared to the reference cross-covariance; the northward-going wavepackets are slowed down by the southward flows. The reverse is true for the southward-going wavepackets. However flows do not result only in a shift of the cross-covariances. Indeed, when we fit the cross-covariances by Gabor wavelets (Equation (\ref{eq.1})), other differences unfold. 

\begin{figure}[ht]
\centering
\includegraphics[width=\textwidth]{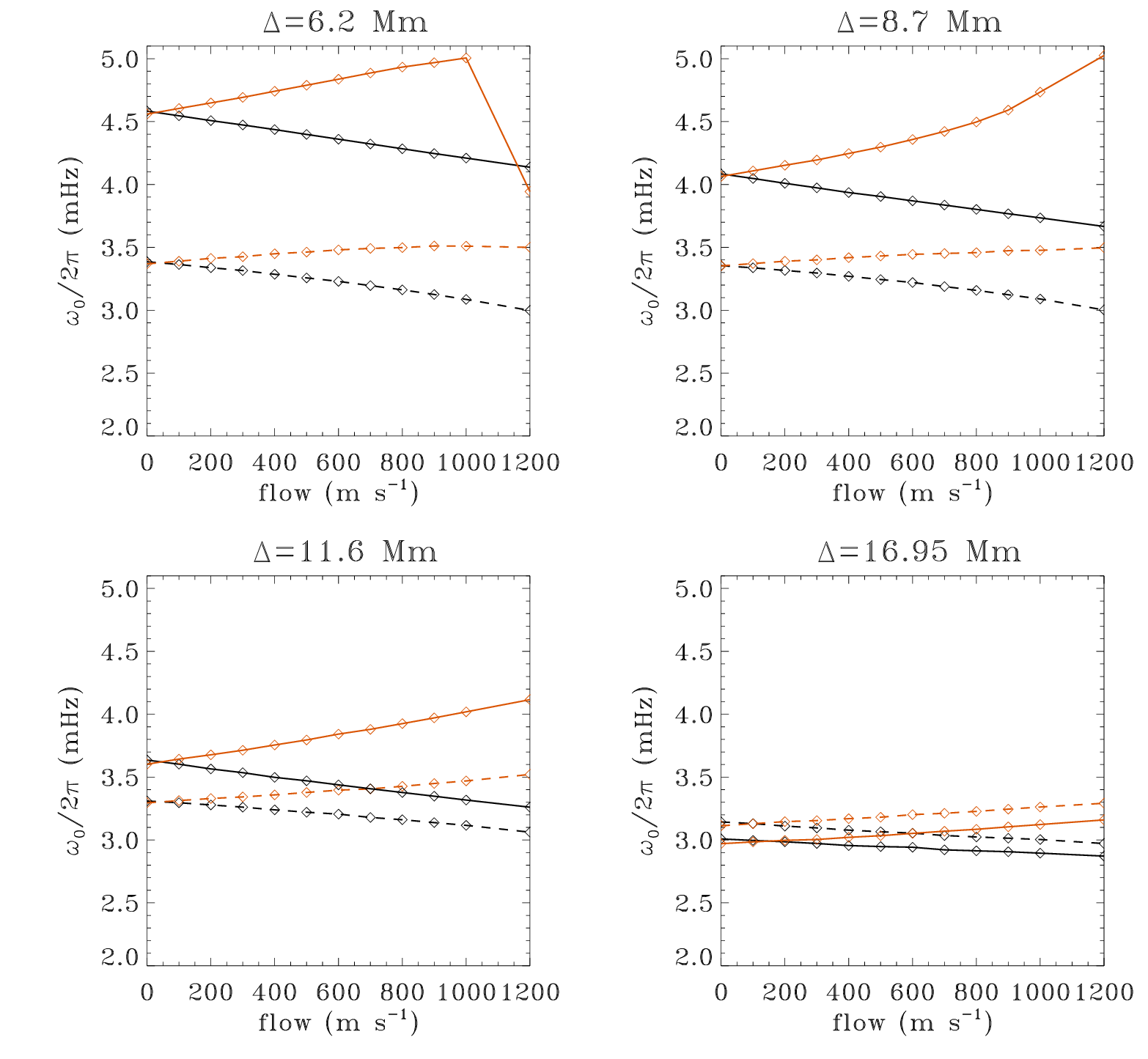}
\caption{Gabor wavelet parameter $\omega_0$ (wavepacket central frequency) as a function of the southward flow velocity. Black lines are for the northward cross-covariances, red lines are for the southward ones. Solid lines are for the standard phase-speed filters, dashed lines are for the broad phase-speed filters. The measurement points are shown as diamonds.}
\label{fig.4} 
\end{figure}

{\it i}) The central frequencies ($\omega_0$) of the wavepackets are also different for the northward and southward cross-covariances (Figure~\ref{fig.4}).
For instance, at $\Delta=8.7$ Mm, in the presence of a $400$ m s$^{-1}$ southward flow, $\omega_0=3.936\times 2\pi$  $10^{-3}$ rad s$^{-1}$ for the spatially-averaged northward cross-covariance, while $\omega_0=4.247\times 2\pi$ $10^{-3}$ rad s$^{-1}$ for the southward cross-covariance.
The change in $\omega_0$  is generally consistent with a Doppler shift $\delta \omega_0$ described by $\delta \omega_0 = \mathbf{k} \cdot \mathbf{u}$ and resulting from a flow $\mathbf{u}$. However at short distances ($\Delta=6.2$ and $8.7$ Mm), and for flows faster than about $u=600$ m s$^{-1}$, $\delta \omega_0$ becomes non-linear in $u$. This non-linearity is probably mainly the result of the application of Gaussian phase-speed filters to the Fourier transform of the data cube. The $\mathbf{k} \cdot \mathbf{u}$ Doppler shift changes the location of the acoustic-mode ridges in regard to the filters and affects the power distribution inside these filters depending on the ridge distance to the filter center. Moreover, this change in the power distribution is different for the positive and negative (in term of $\omega$) branches of the acoustic-mode ridges, explaining why the $\omega_0$ of northward and southward cross-covariances are not symmetrical around the $\omega_0$ value at rest ($u=0$).
With the broad phase-speed filters at $\Delta=6.2$ Mm, the acoustic wavepackets have a lower central frequency ($\omega_0$) and a lower central wavenumber ($k$) where there is more acoustic power than with the standard filters. Therefore, with the broader filters, $\delta \omega_0$ is smaller for a given velocity $u$, because the average $k$ of the wavepacket is smaller.
Finally, $\delta \omega_0$ becomes non-linear in $u$ earlier with the broad filters than with the standard ones, at least at short $\Delta$: this might be because the average $k$ is more dependent on the flow amplitude than in the case of the standard filters.

\begin{figure}[ht]
\centering
\includegraphics[width=\textwidth]{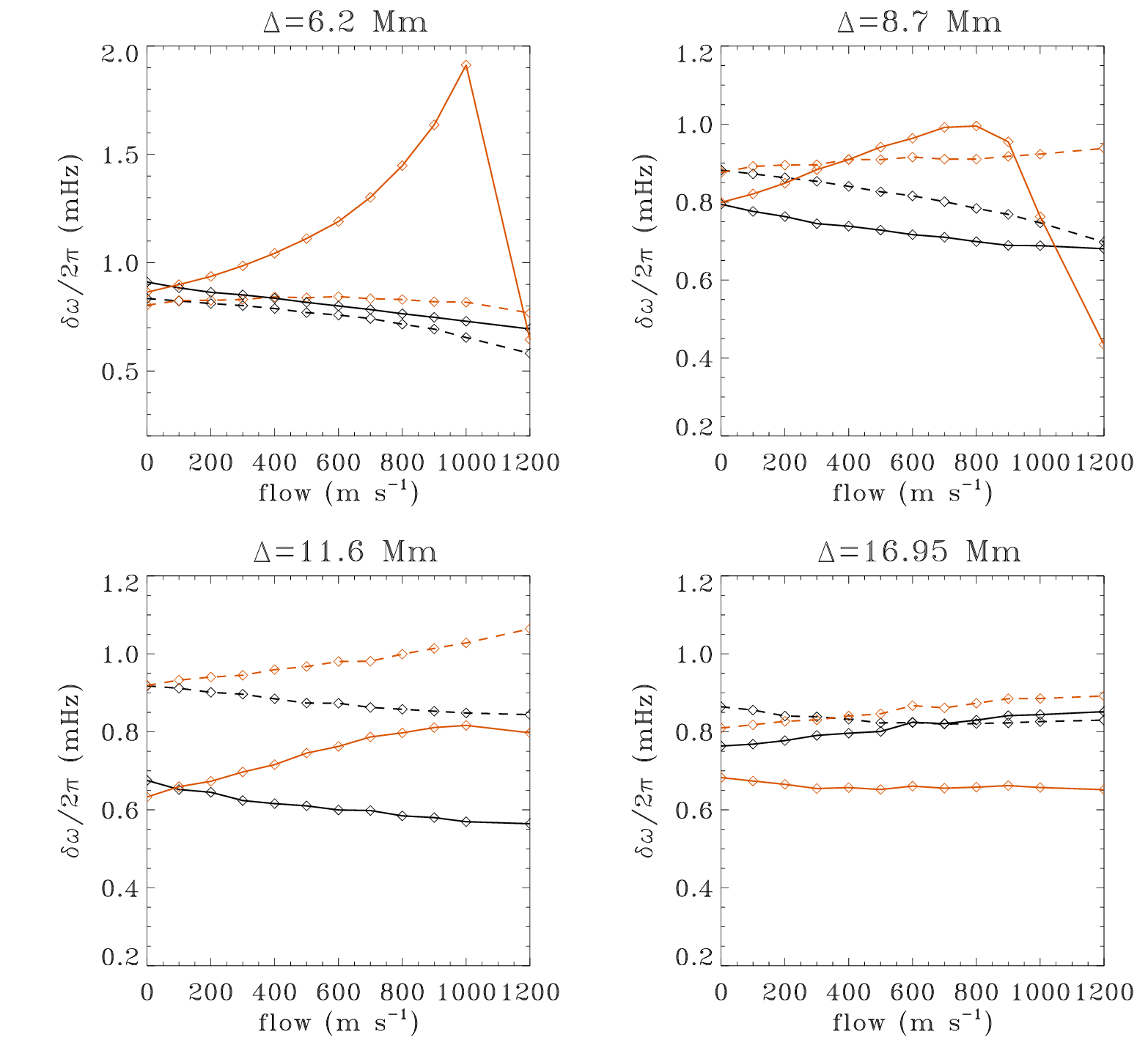}
\caption{Gabor wavelet parameter $\delta \omega$ (wavepacket width) as a function of the southward flow velocity. Black lines are for the northward cross-covariances, red lines are for the southward ones. Solid lines are for the standard phase-speed filters, dashed lines are for the broad phase-speed filters. The measurement points are shown as diamonds.}
\label{fig.5} 
\end{figure}

{\it ii}) The frequency width of the wavepacket varies (Figure \ref{fig.5}): for $u=400$ m s$^{-1}$, $\delta\omega=0.74\times 2\pi$ $10^{-3}$ rad s$^{-1}$ for the northward cross-covariance and $\delta\omega=0.91\times 2\pi$ $10^{-3}$ rad s$^{-1}$ for the southward one.
A change in the wavepacket frequency width is expected because the Doppler shift $\delta \omega_0$ depends on $k$, but as with the change in $\omega_0$, the interplay between the $\omega$-shifted acoustic ridges of the power spectrum and the Gaussian phase-speed filters probably explains the details of this frequency width change.

{\it iii}) The maximum amplitude of the cross-covariances is also affected by the presence of a flow, and again this can partly be explained by a change in the amount of acoustic power selected by the phase-speed filters due to the Doppler shift.

\begin{figure}[ht]
\centering
\includegraphics[height=150mm]{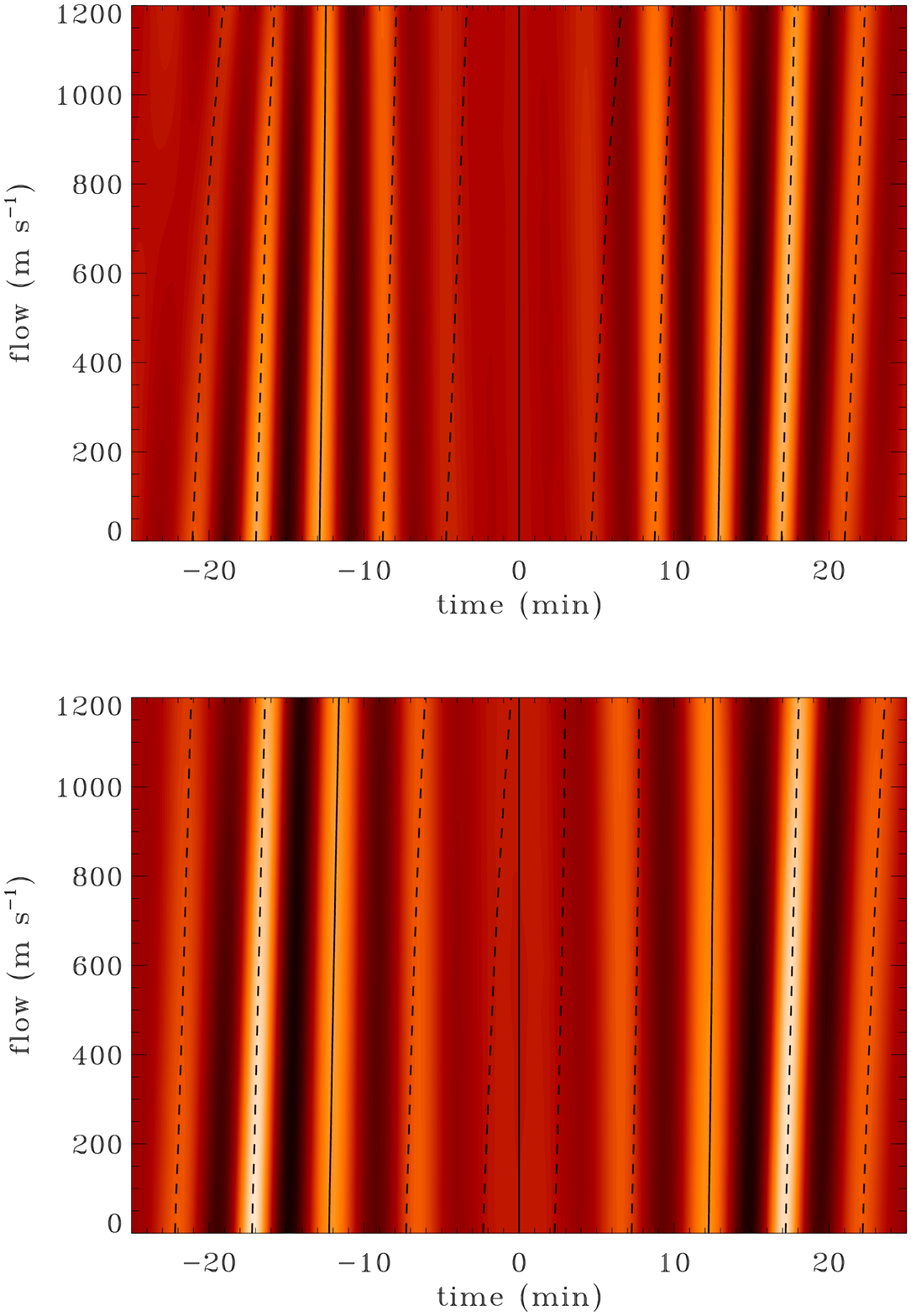}
\caption{Cross-covariances as a function of time and flow amplitude, for a distance $\Delta=8.7$ Mm, and for standard phase-speed filters (upper panel) and broad phase-speed filters (lower panel). Superimposed in black lines are the phase travel times $\taup$ of the Gabor wavelet fit selected for this paper (solid lines) and some other phase travel times $\taup \pm k \times 2\pi/\omega_0$ where $k$ is an integer (dashed lines).
Notice that the filter artifacts at roughly $|t|<10$~min are largely independent of flow amplitude.}
\label{fig.0d}       
\end{figure}

All of these changes affect the northward and southward cross-covariances differently. This can also been seen in Figure~\ref{fig.0d} which shows the cross-covariances as a function of both the time and the flow amplitude for standard and broad phase-speed filters.

There is a clear impact of the FWHM of the filter on the shape of the cross-covariances (and, consequently, on the values of the travel times returned by the definitions studied here). This impact has been mentioned regarding some Gabor wavelet parameters (Figures~\ref{fig.4} and~\ref{fig.5}).
Another consequence of the phase-speed filters, not modeled by the Gabor wavelet, is the presence of a filter artifact (\mbox{e.g.} Couvidat {\it et al.}, 2005).
Indeed, the cross-covariances for the distances studied in this paper have three main components: a filter artifact (at small times $|t|$), a first-bounce skip (waves travelling directly between the two observations points), and a second-bounce skip (waves being reflected once at the surface in between $\br_1$ and $\br_2$).
At $\Delta=6.2$ and $8.7$~Mm, the filter artifact, the first-bounce skip, and the second-bounce skips are entangled. It appears in Figures~\ref{fig.0} and \ref{fig.0d} that the filter artifact is the part of the cross-covariance that is the least affected by the presence of a flow: the location and amplitude of its peaks do not vary much. Standard phase-speed filters engender filter artifacts whose amplitude is larger than those arising from broad filters.
Fast, southward flows push the first-bounce skip of the southward cross-covariances toward lower times $|t|$. Therefore, the peaks of this first-bounce skip become more and more altered by the presence of the filter artifact. 
On the other hand, the use of broad phase-speed filters reduce the amplitude of the artifact and the peaks of the first-bounce skip are less distorted.
An example of the time window [$f(t)$] used to select the first-bounce skip is shown in Figure~\ref{fig.0}. For the travel-time measurements this time-window is always centered at the same time  (independent of the flow amplitude). 

\begin{figure}[ht]
\centering
\includegraphics[width=\textwidth]{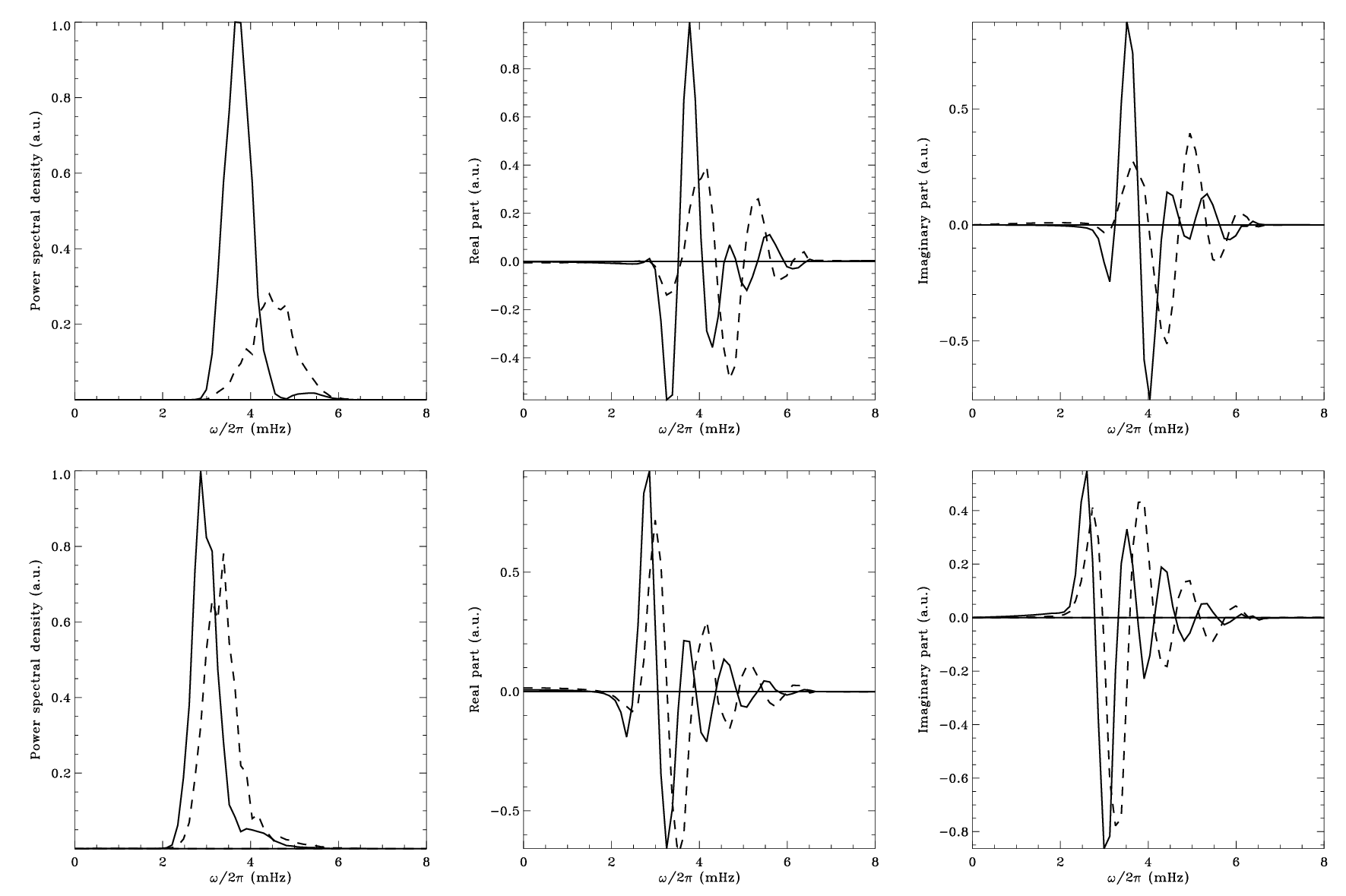}
\caption{Power spectral density (left panels), real (middle panels), and imaginary (right panels) parts of the Fourier transform of the northward (solid lines) and southward (dashed lines) spatially averaged cross-covariances at $\Delta=8.7$ Mm in the presence of a $1000$ m s$^{-1}$ southward flow. Upper panels: standard phase-speed filter; lower panels: broad phase-speed filter.}
\label{fig.0b} 
\end{figure}

Figure \ref{fig.0b} shows the power-spectral density, and the real and imaginary parts of the Fourier transform of the northward and southward spatially averaged cross-covariances in the presence of a $1000$ m s$^{-1}$ southward flow, for the standard and broad phase-speed filters at $\Delta=8.7$ Mm. The differences mentioned earlier in central frequency, frequency width, and total power of the wavepacket between northward and southward cross-covariances are clearly visible. It confirms that the asymmetry between southward and northward cross-covariances is mainly a result of the interaction between the phase-speed filters and the wavepackets. Overall, broader phase-speed filters seem preferable when computing the cross-covariances, since narrow filters introduce more asymmetries between northward and southward cross-covariances and significantly alter the shape of the first-bounce ridge at short distances $\Delta$. This could lead to spurious interpretations regarding physical phenomena like subsurface flows or sound-speed velocity perturbations, when the resulting travel times are inverted with sensitivity kernels that do not take into account the details of the travel-time measurement technique.

\subsection{Definition of the Phase Travel Time With the Gabor Wavelet}

As mentioned in Section $3.2$, when fitting a temporal cross-covariance with a Gabor wavelet we obtain two kinds of travel times: a group $\taug$ and a phase $\taup$ travel time.
A well-known issue with $\taup$ is that it is not uniquely defined ({\it e.g.} D'Silva, 2001). Indeed any time congruent to $\taup$ modulo $2\pi/\omega_0$ is an equally valid phase travel time because the Gabor wavelet is the same for any of these numbers.

\begin{figure}[ht]
\centering
\includegraphics[width=\textwidth]{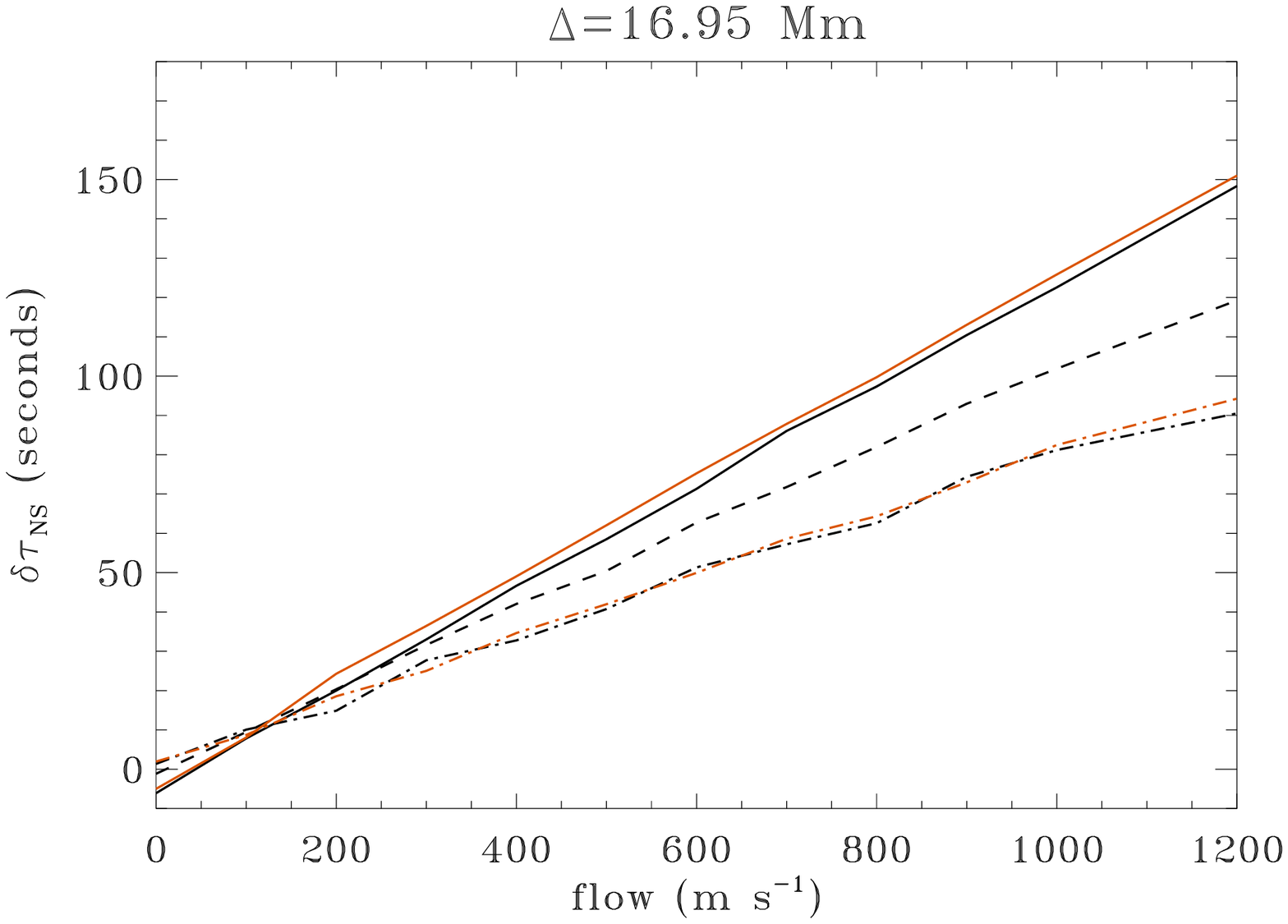}
\caption{Impact of the choice of the Gabor wavelet phase travel time on the relation between North\,--\,South difference travel-time perturbations and the southward flow velocity, at a distance $\Delta=16.95$ Mm. Black lines are for the standard phase-speed filter and red lines are for the broad filter. Solid lines are for a reference $\taup=29.05$ minutes (standard filter) or $29.25$~minutes (broad filter), dashed line is for $\taup=23.5$~minutes, and dash-dotted lines are for $\taup=17.96$~minutes (standard filter) or $18.63$~minutes (broad filter). The group travel time is $\taug=27.69$~minutes (standard filter) or $27.71$~minutes (broad filter).}
\label{fig.0c}       
\end{figure}

In return, the lack of uniqueness of $\taup$ means that the value of $\delta\tau_{\mathrm{NS}}(\br,\Delta)$ engendered by a southward flow is also not unique and depends on the value of the selected reference phase travel time.
Suppose that we select $12.85$~minutes as the phase travel time of $C^{\mathrm{ref}}(\Delta,t)$ for $\Delta=8.7$~Mm, and we add a $200$~m s$^{-1}$ southward flow to the numerical simulation data cube. Then the Gabor wavelet fitting code returns $\tau_{\mathrm{N}}=12.917$~minutes and  $\tau_{\mathrm{S}}=12.781$~minutes, resulting in $\delta\tau_{\mathrm{NS}}=\tau_{\mathrm{NS}}=8.15$~seconds. If we select as reference phase travel time $12.85+2\pi/\omega_0=16.95$ minutes instead, the fitting code returns $\tau_{\mathrm{N}}=12.917+2\pi/\omega_{0;\mathrm{N}}=17.074$~minutes and $\tau_{\mathrm{S}}=12.781+2\pi/\omega_{0;\mathrm{S}}=16.794$~minutes, resulting in $\delta\tau_{\mathrm{NS}}=16.79$~seconds. Indeed, the $\omega_0$ parameter for the northward cross-covariance is $\omega_{0;\mathrm{N}}=4.009\times 2\pi$ $10^{-3}$~rad~s$^{-1}$ while for the southward cross-covariance it is $\omega_{0;\mathrm{S}}=4.153\times 2\pi$ $10^{-3}$~rad~s$^{-1}$. This difference in $\omega_0$ is a direct consequence of the $200$~m s$^{-1}$ flow. This example shows that the same flow velocity can produce different values for $\delta\tau_{\mathrm{NS}}(\br,\Delta)$ (here $8.15$ and $16.79$~seconds). This is a major difference between the three travel-time definitions studied here: GB02 and GB04 return a uniquely defined travel time, unlike the phase travel time of the Gabor wavelet.
Figure \ref{fig.0c} shows the impact of the reference phase travel time on $\delta\tau_{\mathrm{NS}}$ as a function of the southward flow velocity, at $\Delta=16.95$ Mm.

Therefore, when using the Gabor wavelet definition it is necessary to define a criterion regarding which values of the phase travel time are to be selected. It is customary ({\it e.g.} J. Zhao, private communication) to choose as phase travel times those whose value is the closest to --- and if two phase travel times are equally distant, smaller than --- the corresponding group travel times (which is uniquely defined). It can be argued that selecting $\taup$ close to $\taug$ makes sense because that is where the cross-covariance peaks have the largest amplitude and are the furthest away from both the filter artifact and the second-bounce skip, therefore making the determination of their location more accurate.
 At $\Delta=8.7$~Mm the reference group travel time is about equally distant from $12.85$ and $16.95$~minutes. Therefore the phase-travel time that we select is $12.85$~minutes. Trying to fit first-bounce skip peaks located at shorter times is a difficult task as can be seen in Figure~\ref{fig.0d}. In this figure, the dashed lines representing  the phase times do not follow the peaks of the cross-covariances for times $|t|<10$~minutes, confirming that these peaks are mainly filter artifacts.  This criterion also has the advantage that the selected $\taup$ is close to the phase travel time $t_p=\int_\Gamma k_{\rm t}/\omega \, \mathrm{d}s$ ($\Gamma$ is the ray path, $k_{\rm t}$ is the total --- not just horizontal --- wavenumber, and d$s$ is the element of path length) returned by the ray-path approximation at short $\Delta$ distances ($11.3$~minutes at $\Delta=8.7$ Mm, when applied to the solar model S of  Christensen-Dalsgaard {\it et al.}, 1996). At large $\Delta$ this is not the case anymore: for instance at $\Delta=16.95$ Mm, $\taug=26.10$~minutes and $t_p=16.51$~minutes according to the ray-path approximation based on model S, while the equal-to-or-smaller-than criterion implies a reference $\taup=23.51$~minutes. The reference phase travel times we select for the rest of this paper when fitting a Gabor wavelet are listed in the last column of Table \ref{table.1}.

\subsection{Comparison of Definitions of Travel Times}

\begin{figure}[h]
\centering
\includegraphics[width=\textwidth]{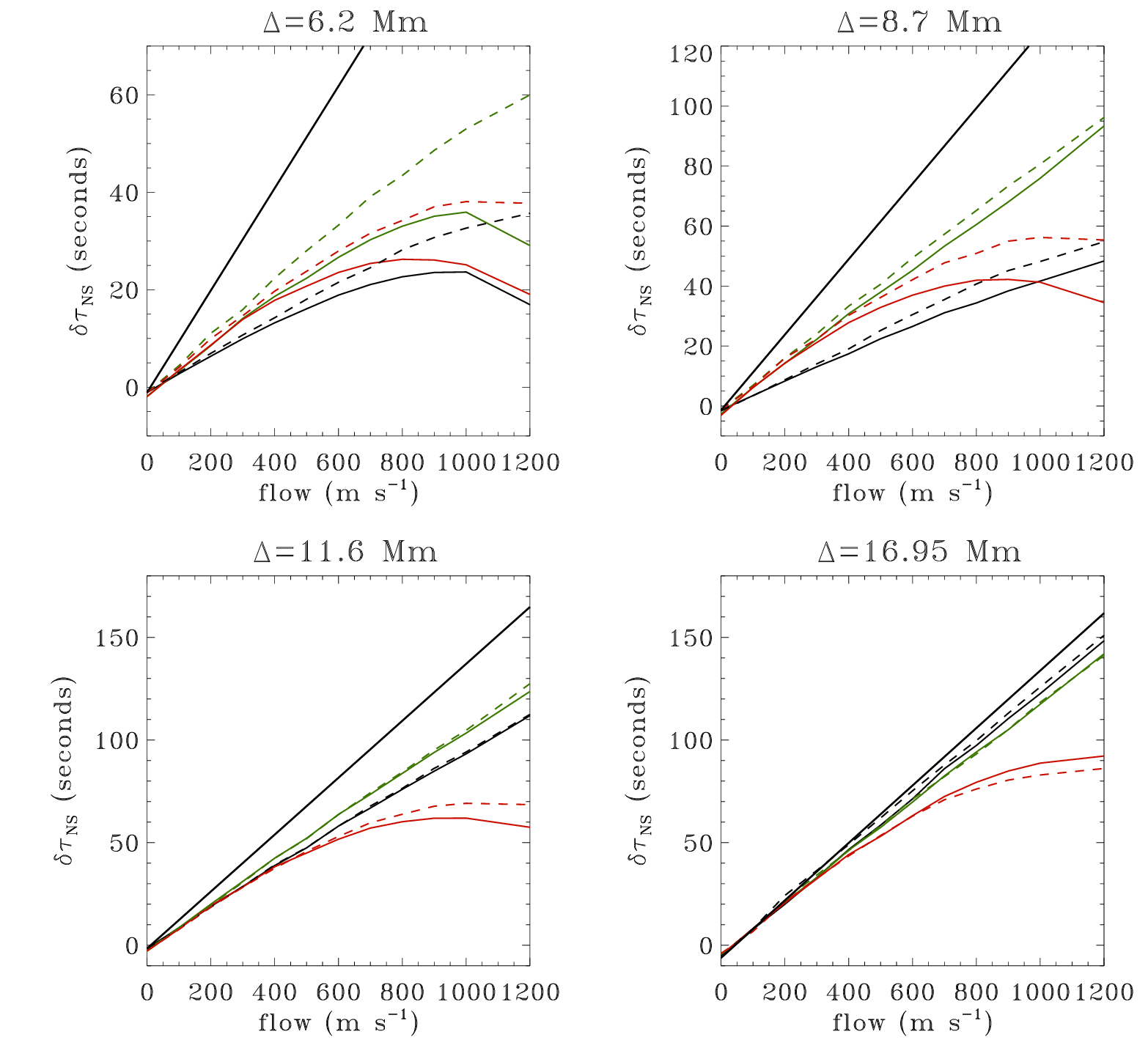}
\caption{North\,--\,South difference travel-time perturbations as a function of the southward flow velocity. The solid lines are for the travel times measured on cross-covariances obtained with the standard width of the phase-speed filters, while the dashed lines are for those obtained with four-times wider phase-speed filters. The black lines are the results of the Gabor wavelet fit, the green lines are for Gizon and Birch (2002), and the red lines are for Gizon and Birch (2004). The perfectly linear curves are the results of the ray-path approximation.}
\label{fig.1}       
\end{figure}

Figure~\ref{fig.1} shows the values of the North\,--\,South difference travel-time perturbation $\delta\tau_{\mathrm{NS}}$ as a function of the amplitude of the horizontal uniform flow. All three travel-time definitions are shown. As mentioned in the previous section, the $\delta\tau_{\mathrm{NS}}$ for the Gabor wavelet fit can be changed by selecting a different reference $\taup$, and thus make them closer to the ones obtained with GB02.
The $\delta\tau_{\mathrm{NS}}(\Delta)$ displayed are the expected values of Gaussian functions fitted to the histograms of $\delta\tau_{\mathrm{NS}}(\br,\Delta)$. Fitting a Gaussian seems more appropriate than reporting the spatial average of $\delta\tau_{\mathrm{NS}}$ because of the presence of numerous misfits with the Gabor wavelet and GB02 definitions (the larger the flow, the more misfits the codes return), and the distribution of $\delta\tau_{\mathrm{NS}}(\br,\Delta)$ at a given $\Delta$ is very close to a Gaussian distribution.
The presence of an intrinsic horizontal flow in the numerical simulation explains why none of the $\delta\tau_{\mathrm{NS}}(\Delta)$ is equal to zero in the absence of any additional flow.
Both the values of $\delta\tau_{\mathrm{NS}}$ measured from cross-covariances obtained with the standard and broad phase-speed filters are shown. 

There is a strong dependence of $\delta\tau_{\mathrm{NS}}$ on the width of the phase-speed filter at small distances $\Delta$. This is hardly surprising considering the significant impact of the phase-speed filters on the shape of the cross-covariances.
Overall, the three travel-time definitions give significantly different answers to the question of what the travel time of a wavepacket is.
At low flow velocities, GB02 and GB04 agree, as expected since GB04 can be seen as a linearization of GB02. However a non-linearity with respect to the flow velocity sets in early for GB04 (at about $200$ m s$^{-1}$ at $\Delta=6.2$ Mm) and the discrepancy with GB02 becomes increasingly larger for larger flows.
Indeed, GB04 appears to be inadequate for the detection of flows larger than about 700 m s$^{-1}$ (at $\Delta=6.2$~Mm and with standard phase-speed filters). We note, however, that GB04 was designed to be applied to quiet-Sun data only. Even though at short $\Delta$ all three travel-time definitions exhibit a non-linearity for the flows tested here, GB04 is the definition departing the earliest from linearity. For the three definitions the non-linearity is larger with narrower filters (except at $\Delta=16.95$~Mm with GB04). This emphasizes the role of the phase-speed filter in the creation of this non-linearity.
At larger $\Delta$, GB02 and the Gabor wavelet fit return similar values for the travel times.

\begin{figure}[h]
\centering
\includegraphics[width=\textwidth]{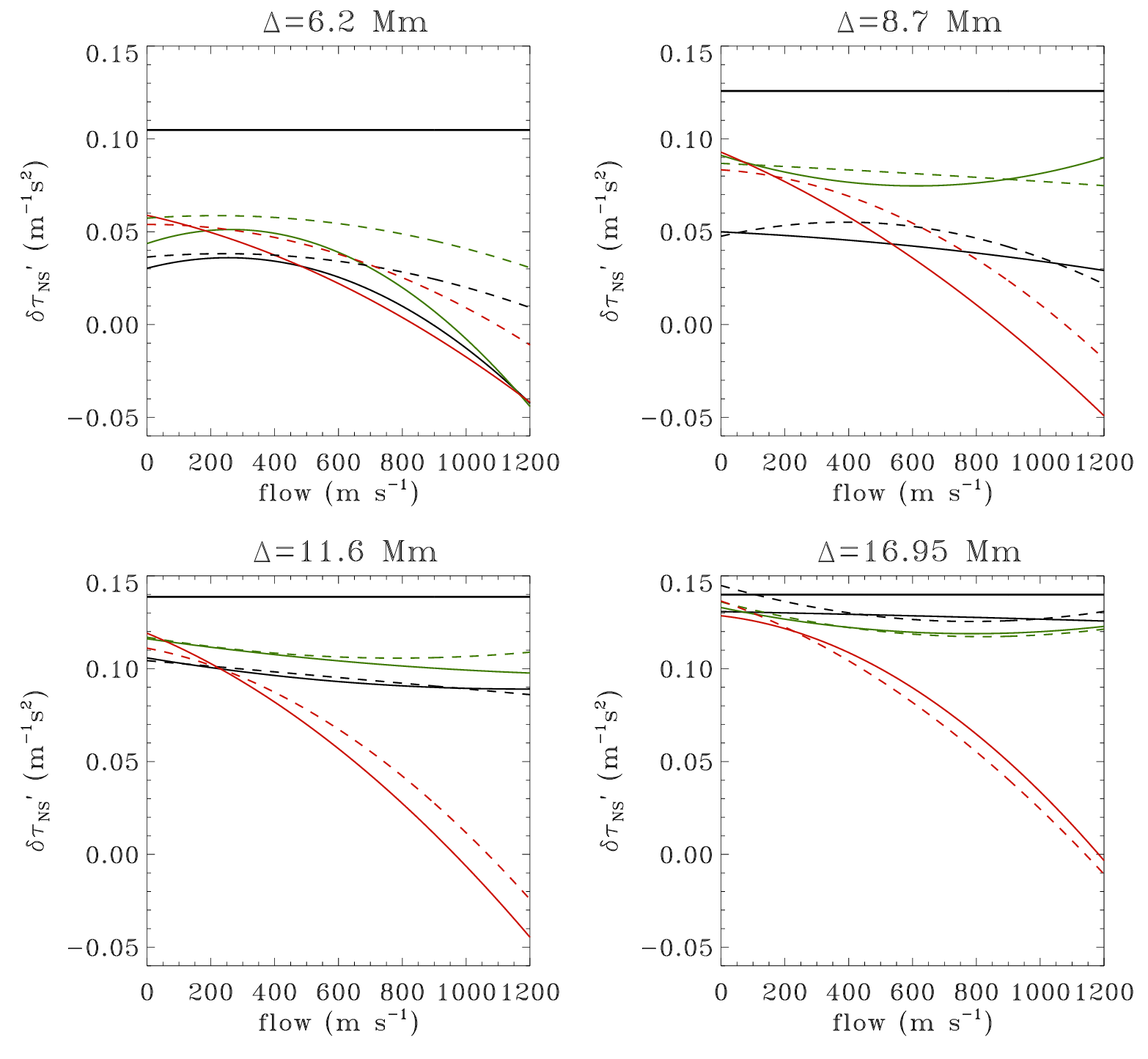}
\caption{Derivatives of the North\,--\,South difference travel-time perturbations with respect to the southward flow velocity. The solid lines are for the travel times measured on cross-covariances obtained with the standard width of the phase-speed filters, while the dashed lines are for those obtained with four-times wider phase-speed filters. The black lines are the results of the Gabor-wavelet fit, the green lines are for Gizon and Birch (2002), and the red lines are for Gizon and Birch (2004). The horizontal solid lines show the results of the ray-path approximation.}
\label{fig.1b}       
\end{figure}

With the standard phase-speed filters, no travel-time definition returns a unique relation between $\delta\tau_{\mathrm{NS}}$ and the flow velocity at $\Delta=6.2$~Mm within the range of flow amplitudes tested. Indeed, a measured $\delta\tau_{\mathrm{NS}}=17$~seconds might be interpreted as arising from a $1200$~m s$^{-1}$ or a $500$~m s$^{-1}$ flow according to the Gabor wavelet fit. This means that the inversion procedure will not be able to detect flows of the order of $1200$ m s$^{-1}$, and will instead assign a $\delta\tau_{\mathrm{NS}}=17$~seconds value to a $500$ m s$^{-1}$ flow, thus underestimating the true flow velocity.
Therefore flows faster than $\approx 1000$~m s$^{-1}$ in the shallow layers underneath the solar surface are impossible to retrieve if we are not careful with the details of the time-distance analysis (width of the phase-speed filters). However such high flow velocities are essentially never present in solar data except in untracked data cubes (cubes not corrected for the solar rotation).
Figure~\ref{fig.1b} shows the derivatives of $\delta\tau_{\mathrm{NS}}$ with respect to the flow amplitude. To display smooth derivatives, the $\delta\tau_{\mathrm{NS}}$ were first fitted with a polynomial of order three and the derivatives were computed from these polynomials. At $\Delta=6.2$~Mm the relation between $\delta\tau_{\mathrm{NS}}$ and flow velocity becomes not linear for flows faster than roughly $300$~m s$^{-1}$ for the Gabor wavelet and GB02 definitions and with standard phase-speed filters. Broader filters delay the onset of non-linearity to about $600-700$ m s$^{-1}$. However, at $\Delta=11.6$ and $16.95$ Mm, GB02 and the Gabor wavelet are approximately linear for the range of flow velocity tested, except maybe at $1200$~m s$^{-1}$. GB04 is never linear in flow velocity over the entire range tested in our study.
At $\Delta=6.2$~Mm, and for standard phase-speed filters, the derivatives of $\delta\tau_{\mathrm{NS}}$ become negative for a ratio of the flow velocity ($v$) to the phase-speed filter central phase speed ($v_0$) larger than $6.5\%$. At this point the definitions of travel times become inadequate because they cannot discriminate between fast and slow flows, as previously mentioned. GB04 is the definition for which the derivatives turn negative at the slowest flows, followed by the Gabor wavelet, and GB02.
The ratio $v/v_0$ at which the derivative of $\delta\tau_{\mathrm{NS}}$ becomes negative becomes smaller and smaller with increasing $\Delta$ for GB04.

The non-linearity and disagreement among travel-time definitions seem to decrease as $\Delta$ increases, which is the opposite of what occurs in Figures 3 and 5 of Jackiewicz {\it et al.} (2007), for a travel-time difference based on GB04. However, Jackiewicz {\it et al.} (2007) used ridge filtering (f-mode filter) and surface-gravity waves, while here we use phase-speed filtering and acoustic waves. Moreover because of differences in the physical models, the oscillation power spectra are also different. Therefore, it is difficult to draw any firm conclusion from this possible discrepancy between the two papers.

Lastly, we remind the reader that only GB04 is linear with respect to the cross-covariance. Therefore, the travel time of the spatially-averaged cross-covariance is not equal to the spatial average of the travel times with GB02 and the Gabor wavelet.  We find that the difference between the average of the travel times and the travel time of the average cross-covariance increases with the flow velocity. This is in part due to the increasing number of misfits we encounter with these two travel-time definitions when the flow velocity increases.

\subsection{Frequency Dependence of the Travel-Time Differences}

\begin{figure}
\centering
\includegraphics[width=\textwidth]{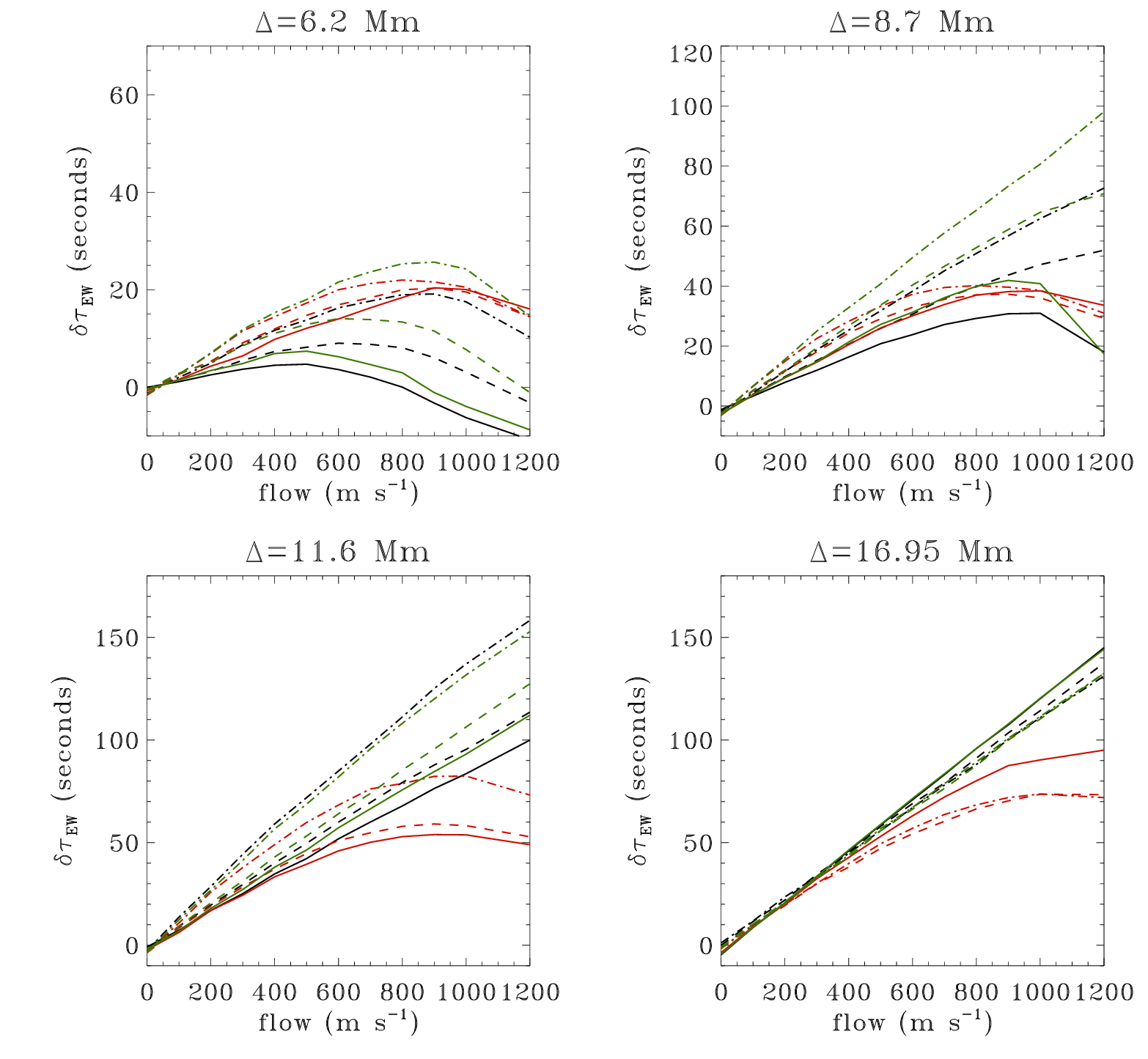}
\caption{North\,--\,South difference travel-time perturbations as a function of the southward flow velocity and when Gaussian frequency filters have been applied on top of the phase-speed filters. All the travel times were obtained with the standard width of the phase-speed filters. The solid lines are for a frequency filter centered on $3.5$~mHz, the dashed lines are for 4~mHz, and the dot-dashed lines are for $4.5$~mHz. The black lines are the results of the Gabor wavelet fit, the green lines are for Gizon and Birch (2002), and the red lines are for Gizon and Birch (2004).}
\label{fig.2}       
\end{figure}

\begin{figure}
\centering
\includegraphics[width=\textwidth]{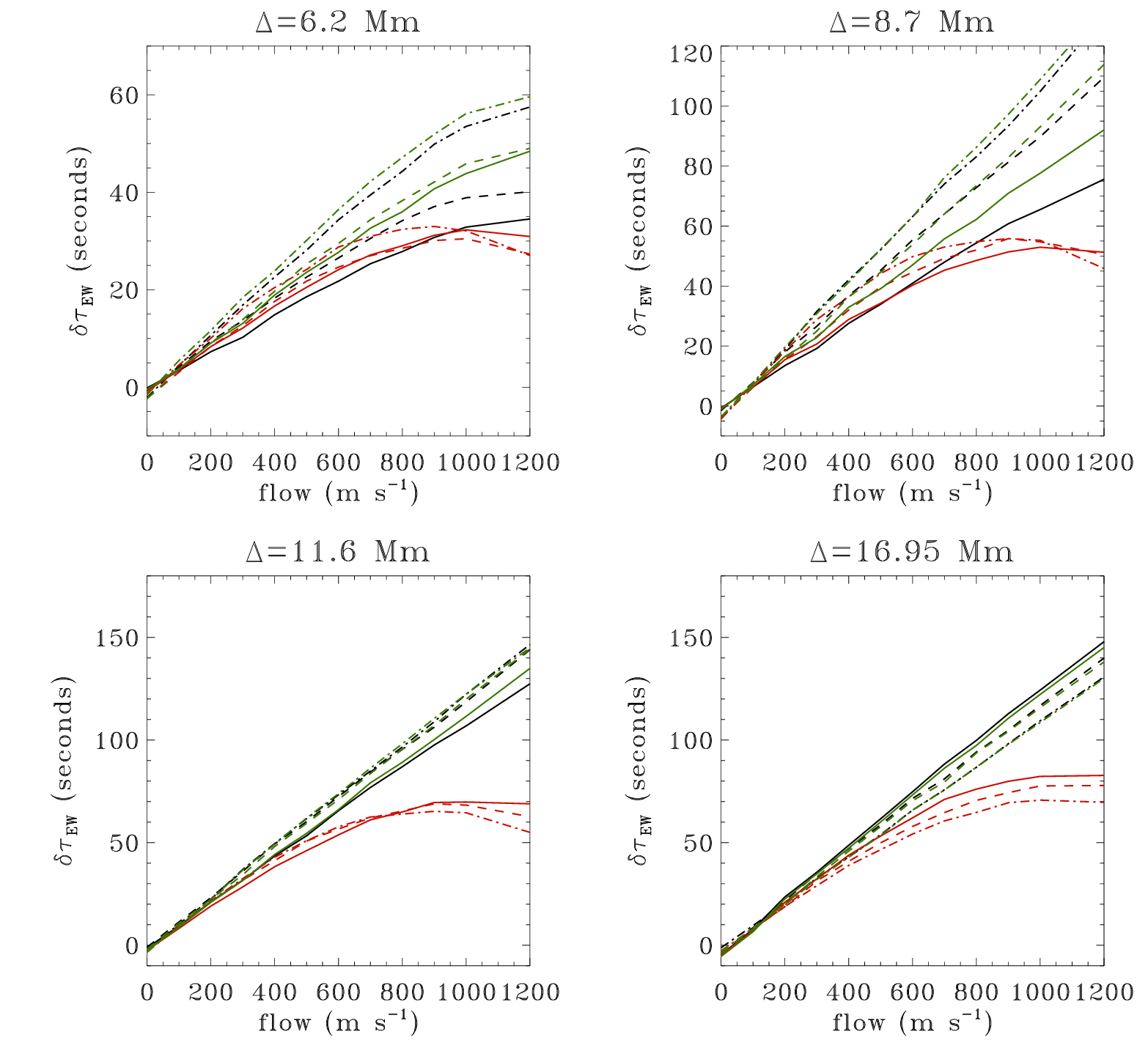}
\caption{North\,--\,South difference travel-time perturbations as a function of the southward flow velocity and when Gaussian frequency filters have been applied on top of the phase-speed filters. All of the travel times were obtained with phase-speed filters four times wider than the standard width. The solid lines are for a frequency filter centered on $3.5$~mHz, the dashed lines are for 4~mHz, and the dot-dashed lines are for $4.5$~mHz. The black lines are the results of the Gabor wavelet fit, the green lines are for Gizon and Birch (2002), and the red lines are for Gizon and Birch (2004).}
\label{fig.3}       
\end{figure}

Following Braun and Birch (2006) and Couvidat and Rajaguru (2007) we applied Gaussian frequency filters to the Fourier transform of the data cube, on top of the phase-speed filters, to select three specific frequency bands.
These three bands are centered on $\nu=3.5$, $4.0$, and $4.5$~mHz. The FWHM of these Gaussian filters is 1~mHz.
Figure~\ref{fig.2} shows the frequency dependence of $\delta \tau_{\mathrm{NS}}$ as a function of flow velocity for cross-covariances obtained with the standard phase-speed filters. Figure~\ref{fig.3} shows the same but for broad phase-speed filters. Again the impact of the phase-speed filter width is clearly visible in these plots. The travel-time difference perturbation exhibits a strong dependence on the frequency. For the Gabor wavelet and GB02 the higher central frequency of the wavepacket leads to larger $\delta \tau_{\mathrm{NS}}$.

We expect to observe a frequency dependence of the travel times: even though the argument that wavepackets with the same central phase-speed ($\omega/k$) follow approximately the same ray path is used to justify the application of phase-speed filters prior to the computation of cross-covariances, the ray theory shows that in actuality there can be a bit of a difference in the trajectories. Thus, both the upper and lower turning points of a wavepacket change with its central frequency. For instance at $\Delta=16.95$ Mm, with the solar model S, the upper turning point is located about $4.8$ km deeper for a wavepacket of central frequency $\nu=3.5$~mHz than for a wavepacket of central frequency $\nu=4.5$~mHz. The lower turning point changes by more.
However the reversal in the sign of $\delta \tau_{\mathrm{NS}}$ observed for fast flows at $\Delta=6.2$~Mm and a frequency of $3.5$~mHz cannot be explained by a small change in the ray-path of the wavepacket. The complex interplay between the phase-speed filter and the wavepacket seems to be the main reason behind this strong non-linearity at low frequency. This result confirms again the interest in broad enough phase-speed filters in time-distance analysis, and that we cannot ignore the details of the travel-time measurement technique when interpreting the travel times.

\subsection{Comparison with the Ray-Path Approximation}

To invert the travel times and retrieve the physical perturbations to which these times are sensitive, sensitivity kernels ({\it i.e.} a model for the linear dependence of travel-time shifts
on perturbations to a solar model) must be employed.
Two main approximations have been used to produce these kernels: the ray-path (Kosovichev and Duvall, 1997) and Born approximations (Birch {\it et al.}, 2001; Gizon and Birch, 2002; Birch and Gizon, 2007, in the context of time-distance helioseismology).
The ray-path approximation (or geometric optics approximation) is valid in the limit of small wavelength while the Born approximation takes into account finite wavelength effects.  In the ray approximation, the travel-time difference $\delta\tau$ caused by a flow $\vel$ is (Kosovichev and Duvall, 1997):
\begin{equation}
\delta\tau = -2\int_\Gamma \frac{\vel\cdot\hat{\bf n}}{c^2} \id s \label{rayp}
\end{equation}
where $\hat{\bf n}$ is the unit vector tangent to the ray path, and $c$ is the sound speed.
In both the ray and Born approximations, the relationship between physical perturbation and travel-time perturbation is assumed to be linear.
As can be seen in Figure \ref{fig.1}, the assumption of linearity with the amplitude of the flow is a major issue when the non-linearity of the travel times sets in. This shows how far the inverted flow velocities could be from the actual flows, for flows with amplitudes above a few hundred m s$^{-1}$. The thick black lines on all four panels show the variation of $\delta \tau_{\mathrm{NS}}(\Delta)$ --- as a function of the flow velocity --- predicted by Equation (\ref{rayp}) applied to model S (which is close to the background model of the numerical simulation). At $\Delta=6.2$ Mm and for standard phase-speed filters, the ray-path kernels seem to overestimate by a factor larger than three the sensitivity of travel times to flows. None of the travel-time definition returns results close to the ones predicted by this ray-path approximation. A way to improve the agreement between predicted and measured travel times with the Gabor wavelet could be to select larger reference phase travel times.
As mentioned in Section $3.1$, our cross-covariances are always averaged over a $\Delta$-distance range to improve their signal-to-noise ratio. On the other hand, the ray-path kernels we compute here are for the specific $\Delta$ listed in Table \ref{table.1} (no average over a range of $\Delta$). Therefore we do not expect perfect agreement between the $\delta \tau_{\mathrm{NS}}(\Delta)$ predicted by the ray-path kernels and the measured $\delta \tau_{\mathrm{NS}}(\Delta)$. However, we also computed cross-covariances that are not averaged over $\Delta$ for comparison, and the travel times that we measured on these cross-covariances are very close to the travel times measured on the averaged cross-covariances (the maximum difference between the two $\delta \tau_{\mathrm{NS}}(\Delta)$ occurred at flows faster than $600$ m s$^{-1}$ and was only about five seconds).

\section{Conclusion}

We added spatially-uniform steady southward flows of amplitudes ranging from 0 to $1200$~m s$^{-1}$ to the vertical velocity data cube of a numerical simulation of solar convection. 
These southward flows produced a perturbation in the North\,--\,South difference travel times of acoustic wavepackets.
We measured these travel times using three different definitions: the Gabor wavelet (Duvall and Kosovichev, 1997), Gizon and Birch (2002), and Gizon and Birch (2004). 
The customary use of the phase travel time returned by the Gabor wavelet fit instead of the group travel time creates an ambiguity regarding the value of the difference travel-time perturbation engendered by a given flow velocity. This is a direct consequence of the lack of uniqueness of the phase travel time together with the fact that horizontal flows shift the central temporal frequency of the wavepackets. This ambiguity in the travel-time difference is a major issue especially when we try to interpret this difference using sensitivity kernels. Conversely, GB02 and GB04 return uniquely defined travel times.
At short distances $\Delta$ between observation points, all three travel-time definitions exhibit a non-linearity with respect to the flow amplitude. At these short distances, the details in the calculation of the cross-covariances, and especially the width of the phase-speed filters, prove crucial.
Broader phase-speed filters may be preferable because they push the onset of non-linearity to larger flow amplitudes.
GB04 is the definition of travel times that departs the earliest from linearity: the onset seems to be at flow amplitudes as low as about $200$~m s$^{-1}$. Therefore the use of GB04 can be problematic when not applied strictly to the quiet Sun. In addition, we note that supergranules have flow velocities that can exceed $200$~m s$^{-1}$, and therefore the use of inversion procedures on travel-time differences obtained by the GB04 definition could return incorrect velocities (in particular underestimated velocities).
GB02 and the Gabor wavelet also exhibit a sudden drop in the travel-time difference at flows faster than 1~km s$^{-1}$ at $\Delta=6.2$ Mm with standard phase-speed filters. This means that fast flows near the solar surface will produce the same travel-time difference as slower flows.  Flows of these amplitude are not common in tracked data cubes where the effect of solar rotation has been largely removed.  The non-linearity of the travel-time shifts is reduced for the case of broad phase-speed filters. 
At distances ($\Delta$) larger than or equal to $16.95$~Mm both GB02 and the Gabor wavelet definitions produce travel-time differences that vary essentially linearly with respect to flow amplitude, within the velocity range studied here. If a non-linearity sets in, it is for flows much faster than $1.2$~km s$^{-1}$. At distances less than about 17~Mm, the sensitivity kernels produced by the ray-path approximation fail to explain the sensitivity of travel-time differences to flows.

\begin{acks}
The work of SC was supported by NASA grants NNX07AK36G (MDI) and NNG05GM85G (LWS). The work of ACB was supported by NASA grant NNH06CD84C. The numerical simulations were performed by Robert F.\ Stein, Ake Nordlund, Dali Georgobiani and David Benson, supported by NASA grants NNG04GB92G and NAG 512450, NSF grants AST-0205500 and AST-0605738, and by grants from the Danish Center for Scientific Computing. The authors thank \mbox{A.G.} Kosovichev for encouraging a work that led to this paper. The authors also thank \mbox{D.} Georgobiani for her help in dealing with the details of the numerical simulation, and \mbox{D.C.} Braun, \mbox{A.D.} Crouch, and \mbox{M.F.} Woodard for useful comments regarding a draft of this paper. 
\end{acks}

\end{article} 
\end{document}